\documentclass[a4paper,twocolumn,11pt,accepted=2022-08-07]{quantumarticle}
\pdfoutput=1
\usepackage[utf8]{inputenc}
\usepackage[english]{babel}
\usepackage[T1]{fontenc}
\usepackage{amsmath}
\usepackage{hyperref}
\usepackage{braket}
\usepackage{tikz}
\usepackage{lipsum}
\newcommand{\tr}{\text{Tr}}

\usepackage[numbers,sort&compress]{natbib}

\begin{document}
	
	\title{Quantifying  fermionic interactions from the violation of Wick's theorem}
	
	\author{Jiannis K. Pachos}
	\affiliation{School of Physics and Astronomy, University of Leeds, Leeds LS2 9JT, United Kingdom}
	\email{J.K.Pachos@leeds.ac.uk}
	\author{Chrysoula Vlachou}
	
	\affiliation{Instituto de Telecomunicações, Av. Rovisco Pais 1, 1049-001 Lisboa, Portugal}
	\affiliation{Departamento de Matemática, Instituto Superior Técnico,
		Universidade de Lisboa, Av. Rovisco Pais 1, 1049-001 Lisboa, Portugal}
\email{chrysoula.vlachou@lx.it.pt}
	\maketitle
	
	\begin{abstract}
	In contrast to interacting systems, the ground state of free systems has a highly ordered pattern of quantum correlations, as witnessed by Wick's decomposition. Here, we quantify the effect of interactions by measuring the violation they cause on Wick's decomposition. In particular, we express this violation in terms of the low entanglement spectrum of fermionic systems. Moreover, we establish a relation between the Wick's theorem violation and the interaction distance, the smallest distance between the reduced density matrix of the system and that of the optimal free model closest to the interacting one. Our work provides the means to quantify the effect of interactions in physical systems though measurable quantum correlations.
	\end{abstract}

\section{Introduction}
\label{sec:intro}

Free fermion systems are trivially integrable and thus are described by an extensive number of conserved quantities. The corresponding conservation laws dictate a particular structure to their energy and entanglement spectra \cite{Haldane,Peschel2}. Moreover, these spectra are given in terms of a number of parameters that grows only polynomially with respect to system size. When interactions are introduced these conserved quantities cease to exist giving rise to energy or entanglement spectra that, in general, are described by an exponential number of parameters \cite{PachosD}. This complexity makes interacting systems hard to investigate and qualitatively understand.

Despite their complexity, interacting systems are responsible for a wide variety of interesting phenomena, such as the fractional quantum Hall effect~\cite{Tsui,Laughlin,QuantumHall}, the emergence of anyonic quasiparticles~\cite{PachosBook,KitaevAnyons}, many-body localisation~\cite{Nandkishore} and quantum many-body scars~\cite{Papic}. Many of these phenomena can be efficiently described in terms of a small number of emerging degrees of freedom. The simplest such scenario is the case where the presence of interactions transforms a system into a free or nearly free system \cite{PachosD}. Identifying the free degrees of freedom enables the efficient description of the system in terms of very few parameters that only grow polynomially with respect to its size. Moreover, the emergence of freedom in interacting systems determines their thermalisation properties, the ballistic/diffusive propagation of quenches and the nature of their quasiparticle excitation~\cite{PachosD}. Surprisingly, there are many interacting systems that, even if they appear to be strongly interacting, they behave effectively as almost free in the thermodynamic limit \cite{AlmostFree}, such as the Ising model in transverse and longitudinal field~\cite{IntDis} or the XYZ model~\cite{Matos}. 

In order to identify the ``freeness" of interacting systems we need a measurable quantity that can reveal if an interacting system is effectively behaving as free, either theoretically or in experiments. To this end several measures have been put forward to identify the emergence of free behaviour~\cite{IntDis}. Here, we propose to employ the violation of Wick's theorem~\cite{Wick} that is written in terms of expectation values \cite{Peschelmonos} which can be in principle measured experimentally, see \cite{ZollerMeas}. If the violation is very small then the system behaves as almost free. The violation of Wick's theorem was previously observed even in non-interacting systems \cite{Wicknonequilibrium}; therein, the authors study the non-equilibrium dynamics of many-body systems, and witness the violation of Wick's theorem due to connected density correlations in the initial state. On the other hand, in our work Wick's violation is employed to provide diagnostics for the ground state of a static system. We also express the violation of Wick's theorem in terms of a few low-entanglement energies of the ground state. This provides a simple physical way to interpret the distribution of the few lowest entanglement spectra of a system  in terms of the applicability of Wick's theorem.

In a different vein, various quantities from quantum information and information geometry have been used in the study of many-body systems in different contexts: from detecting the presence of phase transitions \cite{KitaevQPTent, NikolaFid,SanperaQPTs,VlachouUhlmann,VlachouDQPT} to characterizing  many-body correlations \cite{EntvsCor,KitaevModel,CorrelationsvonNeumann,SanperaReview, Swingle,FMMV} and describing the dynamics of such systems \cite{Calabrese, Moitra, Hamma}.
Along these lines, we also relate the violation of Wick's theorem to the interaction distance~\cite{IntDis}, a quantum information-theoretical quantity that manifests the extent of interactions in the quantum correlations of a system. The interaction distance is the trace distance between the reduced density matrix of the system's ground state from the density matrix of the closest possible free system \cite{StateDiscr}. While the interaction distance is an optimal theoretical way to infer emergent gaussianity, its relation to the Wick's theorem violation provides an intuitive way to understand its properties and to experimentally estimate its value through measurements of quantum correlations (e.g. \cite{GaussianCorExp,localdensboson,highordecorbosons,PHMott}),  as one can relate the operators of the original and the entanglement Hamiltonian \cite{Peschel, Zoller}.  Along these lines, in \cite{Matos} one can find  an example of applying our approach to the XYZ model in the simple case of a system with only two fermionic modes. Here, we go beyond, by presenting also the more physically relevant case of systems with any number of fermionic modes. Moreover, the relation between Wick's violation and the interaction distance was only sketched in the Supplemental Material of \cite{Matos}, while here we present it in full detail and with rigorous proofs. In other words, the current article is the theoretical backbone the underpins the numerical work in \cite{Matos}.

This paper is organised as follows: In Section \ref{sec:freefermions} we consider systems of free fermions. In particular, in Subsection \ref{subsec:expvalspecfreef} we present their entanglement spectra, and with respect to these we calculate the expectation value of the density operator of a fermionic mode. In Subsection \ref{subsec:wickfreef} we present Wick's theorem and derive its form for such systems without interactions. In Section \ref{sec:interf} we consider systems of interacting fermions. In Subsection \ref{subsec:expvalspecinterf}, we employ their entanglement spectra to calculate the expectation values of density operators involving one and two fermionic modes. Furthermore, we define a quantity that evaluates the violation of Wick's theorem in interacting systems, and depends on the expectation values of density operators involving one and two fermionic modes. To illustrate  our method, in Subsection \ref{subsubsec:twomodes} we consider as an example the simple case of a system of interacting fermions with two fermionic modes, while in Subsection \ref{subsubsec:manymodes} we study the general case of a system with $N$ fermionic modes. In Subsection \ref{subsec:wickinterf}, we show that the quantity we defined to indicate the violation of Wick's decomposition  can be bounded from above by the interaction distance. Finally, in Section \ref{sec:conclusions} we present a summary of our results and point out directions of future work.

\section{Free fermions}
\label{sec:freefermions}

We start by studying the behaviour of free fermions. We identify the pattern of quantum correlations exhibited by the ground state of free fermion systems expressed in terms of its entanglement spectrum. In particular, we review the applicability of Wick's theorem for determining  two-point correlation functions. This presentation will help us to subsequently define measures that quantify the deviation from free-spectra patterns when interactions are present. 

\subsection{Expectation values and entanglement spectra of free fermions}
\label{subsec:expvalspecfreef}

Consider a free fermion system in its ground state, $\ket{\psi_0}$, and its bipartition in subsystems $A$ and $B$. The reduced density matrix $\rho =\text{tr}_B(\ket{\psi_0}\!\bra{\psi_0})$ can be expressed as a thermal state 
\begin{equation}
\rho = {e^{-H_E} \over Z}, 
\label{eq:rho}
\end{equation}
where $H_E$ is the entanglement Hamiltonian and $Z$ the corresponding partition function, given as $Z=\tr (e^{-H_E})$. Since the fermionic model is free, its entanglement Hamiltonian is also free~\cite{Peschel}, and can be given in terms of $N$ \emph{fermionic eigenoperators} $a_i$, $a_i^\dagger$ associated to $N$ \emph{fermionic modes}, as 
\begin{equation}
H_E = \sum_{i=1}^N \epsilon_i a^\dagger_i a_i,
\label{eqn:entHam}
\end{equation}
where $\epsilon_i$ are the \emph{single-mode entanglement
	energies} for $i=1,\ldots, N$. Note that we can absorb the partition function $Z$ in $H_E$ by shifting the overall energy by $E_0 \neq 0$, i.e. $H_E=E_0+\sum_{i=1}^N \epsilon_i a^\dagger_i a_i$, as in Ref.~\cite{IntDis}.  
Equivalently, we can write  \eqref{eqn:entHam} in terms of the corresponding \emph{single-mode density operators} $\hat{n}_i=a^\dagger_i a_i$ for $i=1,\ldots, N$, as 
\begin{equation}
H_E = \sum_{i=1}^N \epsilon_i \hat{n}_i.
\label{eqn:entHamn}
\end{equation}
The entanglement spectrum of $H_E$ for a mode $k$ is given by
\begin{equation}
E_k = \sum_{i=1}^N \epsilon_i n_i,
\label{eqn:free}
\end{equation}
where $n_i=0,1$ are the eigenvalues of $\hat{n}_i$. For a density matrix $\rho$ as in  \eqref{eq:rho}, the expectation value of the single-mode density operator $\hat{n}_k$ for some mode $k$ is 
\begin{equation}
\langle \hat{n}_k\rangle_\rho = \tr (\hat{n}_k \rho).
\label{eqn:2point}
\end{equation}
To introduce our notation and techniques employed in later sections, let us explicitly calculate $\langle n_k \rangle_{\rho}$ following well-established steps. Let us start with the partition function for the state  $\rho$ as given in  \eqref{eq:rho}. As the terms of $H_E$ in  \eqref{eqn:entHam} commute with each other we can write the partition function as

	\begin{align}
	Z &= \tr\left(\prod_{i=1}^N e^{-\epsilon_i \hat{n}_i}\right)= \prod_{i=1}^N \left(\sum_{n_i=0}^{1}e^{-\epsilon_in_i}\right)\nonumber\\
	&=\prod_{i=1}^N\left(1+e^{-\epsilon_i }\right),
	\label{en:Z}
	\end{align}
where the trace is calculated with respect to the two possible values of the occupation number, $n_i=0,1$, of each fermionic mode $i$. The expectation value of the single-mode density operator for some mode $k$ becomes
\begin{align}
\langle \hat{n}_k\rangle_\rho&=\tr\left(\hat{n}_k\frac{e^{-H_E}}{Z}\right)=\frac{1}{Z}\tr\left(\hat{n}_ke^{-\sum_{i=1}^{N}\epsilon_i \hat{n}_i}\right)\nonumber\\
&=\frac{1}{Z}\tr\left[-\frac{\partial}{\partial \epsilon_k}\left(e^{-H_E}\right)\right]
  =-\frac{1}{Z}\frac{\partial Z}{\partial \epsilon_k}.
\end{align} 

By employing  \eqref{en:Z}  we obtain
	\begin{equation}
	\langle \hat{n}_k\rangle_\rho 	={1 \over 1+ e^{\epsilon_k}},
	\end{equation} 
i.e., the expectation value of the density for the $k-$th eigenmode  in terms of the single-mode entanglement energy, $\epsilon_k$. 

\subsection{Wick's theorem for free fermions}
\label{subsec:wickfreef}

In the case of free-fermion systems Wick's theorem provides the means to calculate the expectation values of many-mode density operators in terms of the expectation values of fewer-mode density operators. Let $i,j$ be two fermionic modes. Then, the general form of Wick's theorem for the \textit{two-mode density operator}  $\hat{n}_i\hat{n}_j=a_i^\dagger a_i a_j^\dagger a_j $ with respect to the reduced density matrix, $\rho$, of the ground state is  
\begin{widetext}
	\begin{equation}
	\langle a_i^\dagger a_i a_j^\dagger a_j \rangle_\rho = \langle a_i^\dagger a_i \rangle_\rho \langle a_j^\dagger a_j \rangle_\rho- \langle a_i^\dagger a_j^\dagger \rangle_\rho \langle a_i a_j \rangle_\rho+\langle a_i^\dagger a_j \rangle_\rho \langle a_i a_j^\dagger \rangle_\rho.
	\label{eqn:Wicksfull}
	\end{equation}
	\end{widetext}

If we choose  $a_i$ to be the eigenoperators of the entanglement Hamiltonian $H_E$ from  \eqref{eqn:entHam} the above equation simplifies to 
$\langle a_i^\dagger a_i a_j^\dagger a_j \rangle_\rho = \langle a_i^\dagger a_i \rangle_\rho \langle a_j^\dagger a_j \rangle_\rho$,
as the last two expectation values in the RHS of  \eqref{eqn:Wicksfull} are necessarily zero with respect to the ground state of the diagonalised $H_E$. In terms of the corresponding density operators this can be written as
\begin{equation}
\langle \hat{n}_i \hat{n}_j \rangle_\rho = \langle \hat{n}_i \rangle_\rho \langle \hat{n}_j \rangle_\rho.
\label{eqn:Wicksn}
\end{equation}
Note that Wick's theorem when applied to non-interacting systems, it is usually expressed in terms of operators with respect to the original Hamiltonian of the system, while here we apply it to the mode-density operators of the entanglement Hamiltonian. One can show that the density operators of the entanglement Hamiltonian can be expressed in terms of the density operators of the original Hamiltonian (see \cite{Zoller,Peschel}), and thus, verify that the form of Wick's theorem that we use is also valid.
This decomposition of the two-mode density operators in the respective single-mode density operators is a result of the absence of interactions, dictating a trivial pattern of quantum correlations between the two fermionic modes.
	\section{Interacting fermionic systems} 
	\label{sec:interf}
	
	We now investigate the behaviour of quantum correlations for interacting fermionic systems. Initially, we want to express the expectation values of density operators for a single mode and for two modes as a function of the entanglement spectra following the same methodology as in the free case. This will help us determine the violation of Wick's theorem, and employ it, in turn, to quantify the effect of interactions in terms of the interaction distance.
	
	\subsection{Expectation values and entanglement spectra of interacting fermions}
	\label{subsec:expvalspecinterf}
	
	For an interacting fermionic system we expect that in general the entanglement Hamiltonian of its ground state is also interacting. We assume that $H_E^{\text{int}}$ is diagonal in some basis of eigenoperators $a_i$ and $a_i^\dagger$. As $H_E^{\text{int}}$ is diagonal it is necessarily expressed in terms of density operators $\hat n_i = a_i^\dagger a_i$. The simplest first term we can write down is the free term given by \eqref{eqn:entHam}. The simplest interaction is the Coulomb-like  two-mode density operator. More complicated higher order interactions are also possible, but we expect the corresponding correlators to be negligible for generic Hamiltonians. Hence, the simplest diagonal entanglement Hamiltonian, $H^\text{int}_E$, can be expressed as
	\begin{equation}
	H^\text{int}_E = \sum_{i=1}^N \epsilon_i \hat{n}_i +  \sum_{i=1}^{N-1}\sum_{j=2;j>i}^N \epsilon_{ij} \hat{n}_i\hat{n}_j + \dotsm,
	\label{eqn:entHamint1}
	\end{equation}
	where $\epsilon_{ij}$ are the \emph{two-mode entanglement energies} of $H^\text{int}_E$. The ellipsis in  \eqref{eqn:entHamint1} refers to terms comprising more than two modes.  As a result, the entanglement spectrum of $H^\text{int}_E$ is given by
	\begin{equation}
	E_k = \sum_{i=1}^N \epsilon_i n_i +  \sum_{i=1}^{N-1}\sum_{j=2;j>i}^N  \epsilon_{ij} n_in_j.
	\end{equation}
	Note that $\epsilon_{ij}$ contributes to $E_k$ only if both modes $i$ and $j$ are populated, signalling the interaction between them. For weak interactions it is expected that the two-mode  energies $\epsilon_{ij}$ are much smaller than the single-mode energies, i.e. $|\epsilon_{ij}|\ll |\epsilon_i|$.
	
	We can express the expectation value of the single-mode density operator $\hat{n}_k$ as a function of the corresponding single-mode entanglement energy $\epsilon_{k}$, as 
	\begin{align}
	\langle \hat{n}_k\rangle_\rho&=\frac{1}{Z}\tr\left(\hat{n}_ke^{-\sum_{i=1}^{N}\epsilon_i \hat{n}_i-\sum_{i=1}^{N-1}\sum_{j=2;j>i}^N \epsilon_{ij} \hat{n}_i\hat{n}_j}\right)\nonumber\\&=-\frac{1}{Z}\frac{\partial Z}{\partial \epsilon_k}.
	\label{eqn:kmode}
	\end{align} 
	Furthermore, the expectation value of the two-mode density operator  $\hat{n}_k\hat{n}_l$ can be analogously expressed as a partial derivative with respect to the corresponding two-mode entanglement energy $\epsilon_{kl}$, as
	\begin{eqnarray}
	\langle \hat{n}_k\hat{n}_l\rangle_\rho=-\frac{1}{Z}\frac{\partial Z}{\partial \epsilon_{kl}}.
	\label{eqn:klmodes}
	\end{eqnarray}
	For a free system an equivalent expression to \eqref{eqn:klmodes} does not exist, as Wick's theorem given by \eqref{eqn:Wicksn} directly provides the two-mode expectation value in terms of single-mode densities. For the ground state of an interacting system the expectation values of single and two-mode density operators are in general unrelated. Hence, we define the violation of Wick's theorem, ${\cal W}(\rho)$ as 
	\begin{equation}
	{\cal W}(\rho) := |\langle \hat{n}_i \hat{n}_j \rangle_\rho - \langle \hat{n}_i \rangle_\rho \langle \hat{n}_j \rangle_\rho|,
	\label{eqn:W}
	\end{equation}
	in order to quantify the effect interactions have on the ground state quantum correlations of the system.  In the following, we will first determine its value in terms of the entanglement spectrum for the simple entanglement Hamiltonian of two interacting fermionic modes. Second, we will consider a system with any number of fermionic modes and we will employ adequate assumptions to derive approximate expressions for $\langle \hat n_k\rangle_{\rho}$ and $\langle \hat n_k \hat n_l \rangle_{\rho}$, which can be used to evaluate ${\cal W}(\rho)$.
	
	\subsubsection{ The two-mode case}
	\label{subsubsec:twomodes}
	
	Using   \eqref{eqn:kmode} and \eqref{eqn:klmodes} we can calculate the violation of Wick's decomposition as defined in  \eqref{eqn:W} for an interacting system of fermions with any number of modes. Here, we present a simple example  illustrating it for the case where the entanglement Hamiltonian has only two interacting fermionic modes. In this case  \eqref{eqn:entHamint1} becomes 
	\begin{equation}
	H_E^\text{int}=\epsilon_1\hat{n}_1+\epsilon_2\hat{n}_2+\epsilon_{12}\hat{n}_1\hat{n}_2.
	\end{equation}
	The entanglement spectrum of $H_E^\text{int}$ is given by $E_0=0$, $E_1=\epsilon_1$, $E_2 =\epsilon_2$, $E_{12}=\epsilon_1+\epsilon_2+\epsilon_{12}$.
	The partition function is 
	\begin{align}
	Z&=\tr \left(e^{-H_E^\text{int}}\right)=\sum_{n_1=0}^{1}\sum_{n_2=0}^{1}e^{-\epsilon_1n_1-\epsilon_2n_2-\epsilon_{12}n_1n_2}\nonumber\\
	&=1+e^{-\epsilon_{1}}+e^{-\epsilon_{2}}+e^{-\epsilon_{1}-\epsilon_{2}-\epsilon_{12}},\label{eqn:2partition}
	\end{align}
	or, written in terms of the entanglement energies, $Z=1+e^{-E_1}+e^{-E_2}+e^{-E_{12}}$. By employing \eqref{eqn:kmode} we have that the expectation value $\langle \hat{n}_1\rangle_\rho$ is given by
	\begin{align}
	\langle \hat{n}_1\rangle_\rho&=-\frac{1}{Z} \frac{\partial Z}{\partial \epsilon_1}\nonumber\\
	&= \frac{1+e^{E_{12} -E_1}}{ 1 + e^{E_{12} -E_1} + e^{E_{12} -E_2} +e^{E_{12}}},
	\label{eqn:n1}
	\end{align}
	and analogously for $\langle \hat{n}_2\rangle_\rho$. By employing \eqref{eqn:klmodes} we find that the expectation value $\langle \hat{n}_1\hat{n}_2\rangle_\rho$ is given by
	\begin{align}
	\langle \hat{n}_1\hat{n}_2\rangle_\rho & = -\frac{1}{Z}\frac{\partial Z}{\partial\epsilon_{12}}\nonumber\\
	& = \frac{1}{1+e^{E_{12}-E_{1}}+e^{E_{12}-E_{2}}+e^{E_{12}}}.
	\label{eqn:bothmodes}
	\end{align}
	Using \eqref{eqn:n1} and \eqref{eqn:bothmodes} we can determine the violation of Wick's decomposition, ${\cal W}(\rho)$, given by \eqref{eqn:W} as
	\begin{equation}
	{\cal W}(\rho)=\frac{\left|1-e^{E_{12}-E_1-E_2}\right|}{(1+e^{E_{12} - E_{1}}+e^{E_{12} - E_{2}}+e^{E_{12}})^2}.
	\label{eq:Two}
	\end{equation}
	Clearly, for non-interacting systems $E_{12}=E_1+E_2$ which gives ${\cal W}(\rho)=0$. 
	\subsubsection{ The many-mode case}
	\label{subsubsec:manymodes}
	In the general case of a system with $N$ fermionic modes, finding closed formulas for $\langle \hat n_k\rangle_{\rho}$ and $\langle \hat n_k \hat n_l \rangle_{\rho}$ (and, in turn, for ${\cal W}(\rho)$) is a  tedious task. One way to simplify the calculation is to assume that the interactions are weak, i.e. the two mode energies $\epsilon_{ij}$ are much smaller in magnitude than the single-mode energies $\epsilon_{k}$ for any modes $i,j,k\in\{1,2,\ldots, N\}$. We can, then, approximate $\langle \hat n_k\rangle_{\rho}$ and $\langle \hat n_k \hat n_l \rangle_{\rho}$ keeping only terms up to first order in $\epsilon_{ij}$. The single-mode density operator is given by 
	\begin{widetext}
	\begin{equation}
	\langle n_k\rangle_\rho=\frac{e^{-\epsilon_{k}}}{Z}\sum_{n_i=0}^1\sum_{n_j=0}^1\left(\prod_{i=1;i\neq k}^Ne^{-\epsilon_i n_i}\right)\left(1-\sum_{i=1;i\neq k}^{N-1}\sum_{j=2;\{j>i, j\neq k\}}^{N}\epsilon_{ij}n_in_j-\sum_{i=1;i<k}^{N-1}\epsilon_{ik}n_i-\sum_{i=2;i>k}^{N}\epsilon_{ki}n_i \right),\label{eqn:nkmain}
	\end{equation}
	\end{widetext}
	where $Z$ is given by \eqref{eqn:Zsingle}, and the two-mode density operator by
	
	\begin{equation}
	\langle n_kn_l\rangle=\frac{e^{-\epsilon_{k}-\epsilon_{l}}}{Z}\prod_{i=1;i\neq k,l}^N(1+e^{-\epsilon_i}),\label{eqn:nknlmain}
	\end{equation}
	where $Z$ is given by \eqref{eqn:Ztwomode} (for a detailed derivation see Appendix \ref{sec:Appendix}).
	The above expressions can, then, be used along with \eqref{eqn:W} to compute ${\cal W}(\rho)$.
	
	In the following we will relate the violation of Wick's decomposition with the interaction distance, that optimally identifies the interactiveness of a fermionic system. 
	
	\subsection{Violation of Wick's theorem for interacting fermions and its relationship with the interaction distance}
	\label{subsec:wickinterf}
	
	Inspired from quantum information, a  measure to define the effect of interactions on the groundstate correlations of a system is the \emph{interaction distance}, $D_{\cal F}(\rho)$~\cite{IntDis}. The interaction distance measures the distance between the ground state quantum correlations and the closest pattern of correlations of a system of free fermions. To define $D_{\cal F}(\rho)$ we consider the ground state $\ket{\psi}$ and the reduced density matrix $\rho = \tr_B(\ket{\psi}\!\bra{\psi})$ obtained from a bipartition of the system in $A$ and its complement $B$. The interaction distance of a state $\rho$ is then given by
	\begin{equation}
	D_{\mathcal{F}}(\rho)=\min_{\sigma\in\mathcal{F}} \frac{1}{2}\tr|\rho-\sigma|,
	\label{eqn:Df}
	\end{equation}
	where $\frac{1}{2}\tr|\rho-\sigma|$ is the trace distance between two quantum states $\rho$ and $\sigma$. The minimisation is over all  states $\sigma$ in the manifold $\mathcal{F}$ of all possible free density matrices. In the presence of interactions, the spectrum of the entanglement Hamiltonian that determines the quantum correlations between $A$ and $B$, can deviate from the pattern of the spectra of free-fermion systems, given in  \eqref{eqn:free}. Hence, $D_{\mathcal{F}}(\rho)$  quantifies how interacting a state $\rho$ is by means of how  far it is from the closest possible free state. 
	
	Similar to the violation of Wick's decomposition,  the interaction distance can successfully identify if free-fermion behaviour can emerge out of a strongly interacting system by means of the corresponding entanglement spectra ~\cite{IntDis, Meichanetzidis, PachosD,Patrick, Vincent, Matos}. However, there is no direct way to measure $D_{\cal F}$ in the laboratory. On the other hand, the violation of Wick's theorem, ${\cal W}$, is given in terms of expectation values of observables that can, in principle, be determined in an experiment. Below, we relate the violation of Wick's theorem for an interacting system with the interaction distance of its ground state. This relation will facilitate the physical interpretation of the interaction distance as well as its estimation, without the need for the optimisation procedure in the definition procedure required in  \eqref{eqn:Df}.
	
	We now demonstrate that the violation of Wick's decomposition ${\cal W}(\rho)$ is upper bounded by the interaction distance, $D_{\cal F}(\rho)$. To begin with, note that for a free state $\sigma$, we know that $\langle \hat{n}_i \hat{n}_j \rangle_\sigma - \langle \hat{n}_i \rangle_\sigma \langle \hat{n}_j \rangle_\sigma =0$. We then have
	
		\begin{widetext}
	\begin{equation}
	{\cal W}(\rho) = \big|\langle \hat{n}_i \hat{n}_j \rangle_\rho - \langle \hat{n}_i \hat{n}_j \rangle_\sigma -\langle \hat{n}_i \rangle_\rho \langle \hat{n}_j \rangle_\rho + \langle \hat{n}_i \rangle_\sigma \langle \hat{n}_j \rangle_\sigma \big|.
	\end{equation}
	\end{widetext}

	By employing the Cauchy-Schwarz inequality, we have
	\begin{widetext}
	\begin{align}
	{\cal W}(\rho)& \leq \big|\langle \hat{n}_i \hat{n}_j \rangle_\rho - \langle \hat{n}_i \hat{n}_j \rangle_\sigma \big| + \big| \langle \hat{n}_i \rangle_\rho \left(\langle \hat{n}_j \rangle_\rho - \langle \hat{n}_j \rangle_\sigma \right)\big| +\big| \left(\langle \hat{n}_i \rangle_\rho - \langle \hat{n}_i \rangle_\sigma\right)\langle \hat{n}_j \rangle_\sigma \big|
	\nonumber \\
	& \leq  \big| \tr\left[\hat{n}_i \hat{n}_j(\rho-\sigma)\right]\big| + 
	\big|\langle \hat{n}_i \rangle_\rho\big| \cdot \big|\tr \left[\hat{n}_j (\rho-\sigma)\right]\big| +
	\big|\tr\left[\hat{n}_i (\rho-\sigma)\right]\big| \cdot \big|\langle \hat{n}_j\rangle_\sigma \big|.
	\label{eqn:ineq1}
	\end{align}
	\end{widetext}
	Writing  the state $\rho-\sigma$ in its diagonal basis as $\rho-\sigma=\sum_zs_z\ket{s_z}\bra{s_z}$,\footnote{In general, the states $\rho$ and  $\sigma$ are not necessarily diagonal in the same basis. We consider, though, such states, because in what follows we will use the interaction distance, which is the \emph{minimum} distance between the interacting state and the manifold of free states. In other words the state $\sigma$ is the free state \emph{closest} to $\rho$. For this optimization the two states must commute, i.e., they are indeed diagonal in the same basis (for details, see \cite{IntDis} and \cite{PachosD}).} we have that for any mode $k$ 
	\begin{align}
	\tr\left[\hat{n}_k (\rho-\sigma)\right] & = \sum_zs_z\braket{s_z|\hat{n}_k|s_z}\nonumber\\
	& \leq \max_z \braket{s_z|\hat{n}_k|s_z}\sum_z|s_z|\nonumber\\
	& =\|\hat n_k\|\sum_z|s_z|=\|\hat n_k\| \tr |\rho-\sigma|,
	\label{eqn:maxeigen}
	\end{align}
	where we define $ \| \hat n_k\|$ to be the largest eigenvalue of $\hat{n}_k$, i.e. $ \| \hat n_k\| = 1$,  the maximum population of the fermionic mode. \eqref{eqn:maxeigen} holds for any free state $\sigma$ that commutes with $\rho$, thus it also holds for the optimal free state determined by the optimisation procedure in the evaluation of $D_{\mathcal{F}}(\rho)$. Therefore, from  \eqref{eqn:maxeigen} and  \eqref{eqn:Df}  we get
	\begin{equation}
	\tr\left[\hat{n}_k (\rho-\sigma)\right]\leq 2 D_{\mathcal{F}}(\rho),
	\end{equation}
	see also~\cite{Vincent}.  Hence, from  \eqref{eqn:ineq1},\eqref{eqn:maxeigen} and using $|\langle \hat{n}_k\rangle_{\rho,\sigma}|\leq 1$, for all $\sigma$, $\rho$ and $k$, we obtain
	\begin{equation}
	{\cal W}(\rho) \leq 6 D_{\cal F}(\rho).
	\label{eq:DfandW}
	\end{equation}
	Hence, the violation of Wick's decomposition, ${\cal W}(\rho)$, is bounded from above by the interaction distance, $D_{\cal F}(\rho)$. Recent investigations have shown that the numerical values of ${\cal W}(\rho)$ and $D_{\cal F}(\rho)$ are often almost identical~\cite{Matos}. This tight relation provides a practical way to estimate the interaction distance in terms of simple expectation values that can be measured in the laboratory, see for example \cite{ZollerMeas}.
	\section{Conclusions and future work}
	\label{sec:conclusions}
	
	In this paper we investigated systems of interacting fermions and the effect interactions have in their quantum correlations. The coupling of interactions gives only very crude means of a system's ``interactiveness". To overcome this we considered the effect interactions have on the quantum correlations of a system. We analysed the violation of Wick's decomposition, a common tool used in free systems to decompose high-order correlations in terms of low-order ones. Specifically, Wick's theorem can be applied to systems of free fermions to decompose ground state expectation values of two-mode density operators into expectation values of single-mode density operators. For systems of interacting fermions such a  decomposition does not hold. It is exactly the extent of this violation that we used here to quantify the effect of interactions between the fermionic modes. 
	
	Our analytic investigation was carried out in terms of the entanglement spectra, thus offering the possibility to translate our findings to quantum information language.  In particular, we expressed the violation of Wick's decomposition in terms of few low-entanglement spectra that faithfully reproduce the dominant correlations in the system. In addition, we related the violation of Wick's decomposition to the interaction distance, which is an optimal measure of ``interactiveness" in terms of quantum correlations. 
	
	While both, the violation of Wick's decomposition ${\cal W}(\rho)$ and the interaction distance $D_{\cal F}(\rho)$, can be theoretically seen as serving the same purpose, the relationship we derived can be very useful in practice. The violation ${\cal W}(\rho)$ is given in terms of expectation values of observables that can in principle be measured in the laboratory, see for instance \cite{ZollerMeas}. Nevertheless, we do not know if ${\cal W}(\rho)$ is optimal in identifying the ``interactiveness" of a system. On the other hand, $D_{\cal F}(\rho)$ is defined as the optimal measure of ``interactiveness". Nevertheless, it lacks a relation to observables, making it hard to relate it to experiments. Here we establish a tight relation between these two quantities that can facilitate the theoretical and experimental investigation of physically relevant models,  as it was numerically verified in \cite{Matos} for the XYZ model in the simple case of two fermionic modes.
	
	Having linked the violation of Wick's decomposition, ${\cal W}$, to the entanglement energies, one could go further to relate it to other quantities associated to the entanglement spectrum, such as the entanglement entropies \cite{Calabrese,Swingle,Moitra} or alternatives presented in \cite{Hamma}. Along these lines ${\cal W}$ could possibly be useful to infer other features of many-body systems, such as their integrability or the lack of it. Besides the interaction distance, one could try to relate $W$ with different quantities in order to find different -- possibly tighter -- bounds.
	Further theoretical work could consist of applying our approach  to interacting bosons or spin systems. Also, one could use our study to experimentally infer the effect of interactions in the groundstate correlations of fermionic systems. 
\begin{acknowledgements}
	The authors would like to thank A. Deger, A. Hallam, G. Matos, K. Meichanetzidis, Z. Papi\'c and   C. J. Turner  for fruitful discussions. C. V. further acknowledges the hospitality of the Theoretical Physics Research Group at the University of Leeds. J.K.P. acknowledges support from EPSRC Grant No. EP/R020612/1. This project was partially funded by the EPSRC (Grant No. EP/R020612/1). The data that support the findings of this study are available from the authors upon request. 
	C. V. acknowledges support from the Security and Quantum Information Group (SQIG) in Instituto de Telecomunica\c c\~oes, Lisbon. This work is funded by the FCT  (Funda\c c\~ao para a Ci\^encia e a Tecnologia) through national funds  FCT I.P. and, when eligible, by COMPETE 2020 FEDER funds,  under Award UIDB/50008/2020 and the Scientific Employment Stimulus -- Individual Call (CEEC Individual) -- 2020.03274.CEECIND/CP1621/CT0003.
\end{acknowledgements}

	\bibliography{biblio}
\bibliographystyle{plainnat}
	
	\onecolumn
	\appendix
	\section{Appendix}
	\label{sec:Appendix}
To simplify the presentation for a system of interacting fermions that has  many modes, we make the assumption of weak interactions, that is the two-mode energies $\epsilon_{ij}, \forall i,j$ are much smaller in magnitude than the single-mode energies  $\epsilon_i, \forall i$. Recall from the main text that we also assume  that our Hamiltonian only contains terms involving at most two different modes.  We will derive the corresponding expressions keeping only up to first order terms in $\epsilon_{ij}$. 

Consider a system  that has in total $N$ modes. To calculate  $\langle n_k\rangle_\rho $ for any $k$, we first write the Hamiltonian  by isolating the terms involving the $k$-th mode, that is 
\begin{eqnarray}
H^{int}_E=\epsilon_kn_k+\sum_{i=1;i\neq k}^N\epsilon_i n_i+\sum_{i=1;i\neq k}^{N-1}\sum_{j=2;\{ j>i, j\neq k\}}^{N}\epsilon_{ij}n_in_j+\sum_{i=1;i<k}^{N-1}\epsilon_{ik}n_in_k+\sum_{i=2;i>k}^{N}\epsilon_{ki}n_kn_i.
\label{eq:Ham1}
\end{eqnarray}
For weak interactions where $|\epsilon_{ij}| \ll |\epsilon_{k}|,\forall i, j, k $, we can keep terms only up to first order in $\epsilon_{ij}$. We  approximate the partition function $Z$ for $H^{int}_E$ given by \eqref{eq:Ham1} to be 
\begin{small}
\begin{align}
Z=&\sum_{n_i=0}^1\sum_{n_j=0}^1\left(\prod_{i=1;i\neq k}^Ne^{-\epsilon_i n_i}\right)\left(1-\sum_{i=1;i\neq k}^{N-1}\sum_{j=2;\{j>i, j\neq k\}}^{N}\epsilon_{ij}n_in_j\right)
\nonumber\\
&+e^{-\epsilon_k}\sum_{n_i=0}^1\sum_{n_j=0}^1\left(\prod_{i=1;i\neq k}^Ne^{-\epsilon_i n_i}\right)\left(1-\sum_{i=1;i\neq k}^{N-1}\sum_{j=2;\{j>i, j\neq k\}}^{N}\epsilon_{ij}n_in_j-\sum_{i=1;i<k}^{N-1}\epsilon_{ik}n_i-\sum_{i=2;i>k}^{N}\epsilon_{ki}n_i \right),\label{eqn:Zsingle}
\end{align}
\end{small}where the summation over $n_k=0,1$ has already been done, while the summations over the $n_i$ and $n_j$ populations still need to be performed. The expectation value $\langle n_k\rangle_\rho =-\frac{1}{Z}\frac{\partial Z}{\partial\epsilon_{k}}$, where the numerator is expanded to first order in $\epsilon_{ij}$, can then be written as

\begin{align}
\langle n_k\rangle_\rho&=\frac{e^{-\epsilon_{k}}}{Z}\sum_{n_i=0}^1\sum_{n_j=0}^1\left(\prod_{i=1;i\neq k}^Ne^{-\epsilon_i n_i}\right)\nonumber\\
&\times\left(1-\sum_{i=1;i\neq k}^{N-1}\sum_{j=2;\{j>i, j\neq k\}}^{N}\epsilon_{ij}n_in_j-\sum_{i=1;i<k}^{N-1}\epsilon_{ik}n_i-\sum_{i=2;i>k}^{N}\epsilon_{ki}n_i \right),\label{eqn:nkappendix}
\end{align}
where $Z$ is given by \eqref{eqn:Zsingle}.

Analogously, for the  expectation value $\langle n_kn_l\rangle_\rho$ we write the Hamiltonian by isolating the terms involving the $k$-th and $l$-th modes, as 
\begin{align}
H^\text{int}_E&=\epsilon_kn_k+\epsilon_ln_l+\epsilon_{kl}n_kn_l+\sum_{i=1;i\neq k,l}^N\epsilon_i n_i+\sum_{i=1;i\neq k,l}^{N-1}\sum_{j=2;\{j>i, j\neq k,l\}}^{N}\epsilon_{ij}n_in_j\nonumber\\
&+\sum_{i=1;\{i<k, i\neq l\}}^{N-1}\epsilon_{ik}n_in_k+\sum_{i=2 ;\{i>k, i\neq l\}}^{N}\epsilon_{ki}n_kn_i+\sum_{i=1;\{i<l, i\neq k\}}^{N-1}\epsilon_{il}n_in_l+\sum_{i=2 ;\{i>l, i\neq k\}}^{N}\epsilon_{li}n_ln_i.
\end{align}
Note that without loss of generality we assume $k<l$.
Assuming again weak interactions and keeping terms only up to first order in $\epsilon_{ij},\epsilon_{ik},\epsilon_{ki},\epsilon_{il},\epsilon_{li},\epsilon_{kl}$, we  can write the partition function $Z$  as
\begin{align}
Z=&\sum_{n_i=0}^1\sum_{n_j=0}^1
\left[
\left(\prod_{i=1;i\neq k,l}^Ne^{-\epsilon_i n_i}\right)\left(1-\sum_{i=1;i\neq k,l}^{N-1}\sum_{j=2;\{j>i, j\neq k,l\}}^{N}\epsilon_{ij}n_in_j\right)\right.\nonumber\\
&+e^{-\epsilon_k}\left(\prod_{i=1;i\neq k,l}^Ne^{-\epsilon_i n_i}\right)\left(1-\sum_{i=1;i\neq k,l}^{N-1}\sum_{j=2;\{j>i, j\neq k,l\}}^{N}\epsilon_{ij}n_in_j-\sum_{i=1;i<k}^{N-1}\epsilon_{ik}n_i-\sum_{i=2;\{i>k,i\neq l\}}^{N}\epsilon_{ki}n_i\right)\nonumber\\
&+e^{-\epsilon_l}\left(\prod_{i=1;i\neq k,l}^Ne^{-\epsilon_i n_i}\right)\left(1-\sum_{i=1;i\neq k,l}^{N-1}\sum_{j=2;\{j>i, j\neq k,l\}}^{N}\epsilon_{ij}n_in_j-\sum_{i=1;\{i<l,i\neq k\}}^{N-1}\epsilon_{il}n_in_l-\sum_{i=2;i>l}^{N}\epsilon_{li}n_ln_i\right)															\nonumber\\
&+e^{-\epsilon_k-\epsilon_l}\left(\prod_{i=1;i\neq k,l}^Ne^{-\epsilon_i n_i}\right)\left(1-\epsilon_{kl}-\sum_{i=1;i\neq k,l}^{N-1}\sum_{j=2;\{j>i, j\neq k,l\}}^{N}\epsilon_{ij}n_in_j -\sum_{i=1;i<k}^{N-1}\epsilon_{ik}n_i\right.
\nonumber\\
&\left.\left.-\sum_{i=2;\{i>k,i\neq l\}}^{N}\epsilon_{ki}n_i-\sum_{i=1;\{i<l,i\neq k\}}^{N-1}\epsilon_{il}n_i-\sum_{i=2;i>l}^{N}\epsilon_{li}n_i \right)\right],
\label{eqn:Ztwomode}
\end{align}
where the summations over $n_k,n_l=0,1$ have been done already, while the summations over the $n_i$ and $n_j$ populations still need to be performed. The expectation value $\langle n_kn_l\rangle =-\frac{1}{Z}\frac{\partial Z}{\partial\epsilon_{kl}}$, where the numerator is expanded to first order in $\epsilon_{ij},\epsilon_{ik},\epsilon_{ki},\epsilon_{il},\epsilon_{li},\epsilon_{kl}$ takes the rather simple form
\begin{eqnarray}
\langle n_kn_l\rangle=\frac{e^{-\epsilon_{k}-\epsilon_{l}}}{Z}\prod_{i=1;i\neq k,l}^N(1+e^{-\epsilon_i}),\label{eqn:nknlappendix}
\end{eqnarray}
where $Z$ is given by \eqref{eqn:Ztwomode}. Using \eqref{eqn:nkappendix} and \eqref{eqn:nknlappendix} we can derive the expressions for the violation of Wick's decomposition $\mathcal{W}(\rho)$, given by \eqref{eqn:W}, for systems with many modes. 

\end{document}